\documentclass[aps,nofootinbib,onecolumn,preprint]{revtex4}%
\usepackage{amsmath}
\usepackage{graphicx}
\usepackage{amsfonts}
\usepackage{amssymb}
\usepackage{array}%
\providecommand{\U}[1]{\protect \rule{.1in}{.1in}}
\begin{document}
\title{Lepton Mixing, Residual Symmetries, and Trigonometric Diophantine Equations}
\author{Bo Hu}
\email{bohu@ncu.edu.cn}
\affiliation{Department of Physics, Nanchang University, Nanchang 330031, China}

\begin{abstract}
In this paper, we study residual symmetries in the lepton sector. Our first
concern is the symmetry of the charged lepton mass matrix in the basis where
the Majorana neutrino mass matrix is diagonal, which is strongly constrained
by the requirement that the symmetry group generated by residual symmetries is
finite. In a recent work R.~M.~Fonseca and W.~Grimus found that there exists a
set of constraint equations that can be completely solved, which is essential
in their approach to the classification of lepton mixing matrices that are
fully determined by residual symmetries. In this paper, a method to handle
trigonometric Diophantine equations is introduced. We will show that the
constraint equations found by Fonseca and Grimus can also be solved by this
method. Detailed derivation and discussion will be presented in a
self-contained way. In addition, we will also show that, in the case where
residual symmetries satisfy a reality condition, this method can be used to
solve the equation constraining parameters in the symmetry assignment that
controls the group structure generated by residual symmetries and is directly
related to mixing matrix elements.

\end{abstract}
\maketitle

\section{Introduction}

In recent years the relation between lepton mixing and residual symmetries
originating from finite flavor groups has received a fair amount of attention
\cite{Lam1, Lam2, grimus3, TFH1, TFH2, Sim1, Sim2, BH, grimus1, Lam5, Sim4,
grimus2}. Other aspects of residual symmetries have also been studied
\cite{rs}. Although the idea that lepton mixing may originate from an
underlying discrete symmetry has been discussed extensively in literature (for
some recent reviews, see, e.g., \cite{fsr}), often only specific residual
symmetries are involved, e.g., in lepton flavor models (see, e.g., \cite{A45,
Delta36, S34, othermodels} for models based on various discrete groups
including $\mathbf{A}_{4}$, $\mathbf{S}_{4}$, $\mathbf{\Delta}(6n^{2})$,
etc.). With the help of the computer algebra system GAP \cite{GAP}, scans of
finite groups have also been performed by several groups to look for viable
discrete flavor symmetries \cite{GScan}. Hence, it is worthwhile to see
whether there are any general and model-independent results that can be
obtained for residual symmetries and the corresponding flavor symmetries and
mixing patterns.

In \cite{Sim1} and \cite{Sim2} Hernandez and Smirnov showed that, if residual
symmetries generate a finite group, then the mixing matrix elements can be
related to parameters in the symmetry assignment, including several
characteristic parameters of residual symmetries. In \cite{BH}, this method
was used to construct full-mixing matrices, i.e., the mixing matrices that are
fully determined by residual symmetries. In particular, using an algebraic
method to solve the unitary condition which must be satisfied by any symmetry
assignment that can generate a full-mixing matrix, we found that all the
full-mixing patterns can be determined in the minimal scenario where residual
symmetries satisfy a reality condition. In more general cases, explicit
expressions for the mixing matrix elements can be obtained (see, e.g.,
\cite{Sim2} and \cite{BH}) and, to obtain specific solutions, algebraic
methods can still be useful, but the problem of finding the complete set of
solutions becomes more involved. Nevertheless, a clear and complete answer to
this problem certainly can improve our understanding of the role that residual
symmetries might play.

Remarkably, a complete classification of lepton mixing matrices from residual
symmetries was obtained by Fonseca and Grimus in a recent work \cite{grimus2}.
Their approach is also based on the assumption that residual symmetries are
originated from a finite flavor group. Nevertheless, in this approach, mixing
matrices are not obtained from symmetry assignments, but from $T$, the
generator of the residual symmetry in the charged lepton sector. Therefore, it
is crucial to find explicitly all the possible forms that $T$ can assume. In
\cite{grimus2}, this is done by first establishing the constraint equations of
$|T_{ij}|$, and then showing that these equations can be solved completely by
using some mathematical results concerning roots of unity.

In this paper, we will show that those constraint equations of $|T_{ij}|$ can
also be solved by a method developed in section III. Briefly speaking, our
method is based on the observation that the equations to be solved can be
written as trigonometric Diophantine equations which can be transformed to
arithmetic equations involving only rational integers. They are not
equivalent, but the latter can be used as auxiliary conditions which may lead
to great simplification. The transformation formula can be written in a simple
form and will be derived in a self-contained way. Some mathematical details
are provided in the appendix.

The solutions to the constraint equations of $|T_{ij}|$ are derived in Sec.~IV
by using the method introduced in section III. We will show that these
equations admit only a few solutions from which the basic forms of $|T|$ can
be obtained. Another use of trigonometric algebraic numbers will also be
discussed. In Sec.~V, we will show that this method can also be used to find
the complete solution of the unitary condition in the minimal scenario. The
results were already given and discussed in \cite{BH}, but the derivation was
omitted in order to avoid being distracted by mathematical
details\footnote{Some discussions in this paper can also be found in the first
arXiv version of \cite{BH}.} and hence the derivation presented in Sec.~V can
complete the discussion of the minimal scenario in \cite{BH}. We hope that the
discussions in this paper and those in other works including \cite{BH},
\cite{grimus1}, and \cite{grimus2} may draw attention to some algebraic
notions and techniques that are much less well-known than group theoretical
techniques but may also be useful in the discussions of discrete flavor symmetries.

Before proceeding to the main discussion, to provide the necessary background
and set our notations, in Sec.~II we briefly review and discuss the two
approaches mentioned above.

\section{Lepton mixing and residual symmetries}

We denote the generator of the symmetry of the left-handed charged lepton mass
matrix $M_{l}M_{l}^{\dag}$ by $T$ and those of the neutrino mass matrix
$M_{\nu}$ by $S_{i}$ where $i=1$, $2$, or $3$. The symmetry groups generated
by $T$ and $S_{i}$ are denoted by $G_{e}$ and $G_{\nu}$, respectively. As
usual we adopt the assumption that neutrinos are Majorana particles and
$G_{\nu}$ is the Klein four-group:%
\[
G_{\nu}=\left \{  \mathbf{1},S_{1},S_{2},S_{3}\right \}  .
\]
In the basis where $M_{\nu}$ is diagonal, $S_{i}$ can be written as%
\begin{equation}
S_{1}^{d}=\mathrm{diag}\{1,-1,-1\},\quad S_{2}^{d}=\mathrm{diag}%
\{-1,1,-1\},\quad S_{3}^{d}=S_{1}^{d}S_{2}^{d}. \label{si}%
\end{equation}
On the other hand, in the basis where $M_{l}$ is diagonal, $T$ becomes
diagonal and will be denoted by $T_{d}$.

As emphasized in the introduction, the following discussion is based on the
assumption that $T$ and $S_{i}$ together generate a finite group $G$, which
may be identified with the flavor group $G_{f}$. A necessary condition for
this assumption to hold is given by the relations \cite{Sim1}%
\begin{equation}
S_{i}^{2}=T^{m}=\left(  W_{i}\right)  ^{p_{i}}=\mathbb{I,} \label{ba}%
\end{equation}
where $m$ and $p_{i}$ are integers and $W_{i}\equiv(S_{i}T)^{-1}$. Since the
lepton mixing matrix $U=U_{l}^{\dag}U_{\nu}$ where $U_{l}$ and $U_{\nu}$ are
the matrices diagonalizing $T$ and $S_{i}$, then from Eq.~(\ref{ba}) one finds
that $|U_{\rho i}|$ can be determined by parameters in the so-called symmetry
assignment including $m$ and $p_{i}$. To see that, let us begin with
$\mathrm{Tr}[W_{i}]$. From the definition of $W_{i}$ it follows that%
\begin{equation}
\left(  \mathrm{Tr}[W_{i}]\right)  ^{\ast}=\mathrm{Tr}[S_{i}T]=\mathrm{Tr}%
[US_{i}^{d}U^{\dag}T_{d}]=2\mathrm{Tr}\left[  A_{i}T_{d}\right]
-\mathrm{Tr}[T_{d}] \label{wia}%
\end{equation}
where%
\[
A_{i}=\mathrm{diag}\{|U_{1i}|^{2},|U_{2i}|^{2},|U_{3i}|^{2}\}.
\]
In general, because $T^{m}=\left(  W_{i}\right)  ^{p_{i}}=\mathbb{I}$, one can
write
\begin{equation}
T_{d}=\mathrm{diag}\{e^{i\left.  2\pi k_{1}\right/  m},e^{i\left.  2\pi
k_{2}\right/  m},e^{i\left.  2\pi k_{3}\right/  m}\},\quad \mathrm{Tr}%
[W_{i}]=\sum_{j=1}^{3}e^{i2\pi n_{j}/p_{i}}. \label{wisum}%
\end{equation}
Then from Eqs.~(\ref{wia}) and (\ref{wisum}) the relations between $|U_{\rho
i}|$ and the parameters in $T_{d}$ and $\mathrm{Tr}[W_{i}]$ can be
established. More details including the explicit expressions can be found in
\cite{Sim1, Sim2, BH} and will not be repeated here except for the minimal
scenario discussed in Sec.~V.

To obtain full-mixing matrices, we shall require that all $S_{i}\in G_{f}$
because one $S_{i}$ fixes only a column in $|U|$. Then, from Eq.~(\ref{wia})
and $\sum_{i=1}^{3}S_{i}^{d}=-\mathbb{I}$, it follows that%
\begin{equation}
\sum_{i=1}^{3}\left(  \mathrm{Tr}[W_{i}]\right)  ^{\ast}=-\mathrm{Tr}[T_{d}].
\label{ucon0}%
\end{equation}
This equation will be referred to as unitarity condition \cite{BH}, which must
be obeyed by any combination of $S_{i}$ and $T$ that generates a full-mixing
matrix\footnote{Note that, if $T$ has degenerate eigenvalues, one will need
another matrix $T^{\prime}$ to determine $|U|$ fully because only one row in
$|U|$ can be obtained from Eq.~(\ref{wia}) even if all $S_{i}\in G_{f}$.}.

As mentioned in the introduction, in \cite{BH} attempts were made to find
possible full-mixing matrices by solving the unitarity condition given by
Eq.~(\ref{ucon0}). It was found that in the minimal scenario where
$\mathrm{Tr}[T_{\alpha}]$ and $\mathrm{Tr}[W_{i}]$ are real, Eq.~(\ref{ucon0})
can be solved completely and then all the possible mixing patterns can also be
obtained from its solutions. The detailed derivation omitted in \cite{BH} can
be found in section V. In non-minimal scenarios, the unitary condition becomes
more involved but less restrictive. Nevertheless, solutions to
Eq.~(\ref{ucon0}) are still severely constrained by its algebraic nature, as
discussed in \cite{BH}, which is also implied by the results of numerical
calculations. Therefore, it is reasonable to think that it might still be
possible to obtain a complete picture of full-mixing patterns under
assumptions mentioned above, although solving Eq.~(\ref{ucon0}) alone may not
be an efficient way because the relations in Eq.~(\ref{ba}) are not sufficient
for $S_{i}$ and $T$ to generate a finite group.

This problem is solved by Fonseca and Grimus in a recent work. In
\cite{grimus2}, they begin with $Y^{(ij)}=T^{\dag}S_{i}TS_{j}$ which must also
have finite orders. Since one can show that $\det Y^{(ij)}=1$ and
$\mathrm{Tr}\left[  Y^{(ij)}\right]  $ are real, it then follows that
$Y^{(ij)}$ have eigenvalues $1$, $\lambda^{(ij)}$, $\left(  \lambda
^{(ij)}\right)  ^{\ast}$, where $\lambda^{(ij)}$ are roots of unity and hence
can be written as%
\begin{equation}
\lambda^{(ij)}=e^{i2\pi \left.  k^{(ij)}\right/  m^{(ij)}} \label{lambdaij}%
\end{equation}
where $k^{(ij)}$ and $m^{(ij)}$ are coprime. Because $\mathrm{Tr}\left[
Y^{(ij)}\right]  $ are basis independent, from $\sum_{i=1}^{3}S_{i}%
^{d}=-\mathbb{I}$, one finds that%
\[
\sum_{i=1}^{3}\mathrm{Tr}\left[  Y^{(ij)}\right]  =\sum_{j=1}^{3}%
\mathrm{Tr}\left[  Y^{(ij)}\right]  =1
\]
which can be written as%
\begin{equation}%
\begin{array}
[c]{l}%
\displaystyle \sum_{i=1}^{3}\left(  \lambda^{(ij)}+\left(  \lambda
^{(ij)}\right)  ^{\ast}\right)  =-2,\\
\displaystyle \sum_{j=1}^{3}\left(  \lambda^{(ij)}+\left(  \lambda
^{(ij)}\right)  ^{\ast}\right)  =-2.
\end{array}
\label{FG3}%
\end{equation}
In addition, in the basis where $S_{i}$ are diagonal, one can show that%
\begin{equation}
4\left \vert T_{ij}\right \vert ^{2}=\mathrm{Tr}\left[  Y^{(ij)}\right]
+1=\lambda^{(ij)}+\left(  \lambda^{(ij)}\right)  ^{\ast}+2. \label{TijAbs2}%
\end{equation}
Therefore, from $\lambda^{(ij)}$ satisfying (\ref{FG3}) one can construct the
matrix $|T|$ which is defined by $|T|_{ij}=\left \vert T_{ij}\right \vert $. It
was found in \cite{grimus2} that there exist only five basic forms that $|T|$
can assume. From these basic forms one can derive all the possible full-mixing
patterns, because in the basis where $S_{i}$ are diagonal the lepton mixing
matrix can be obtained from $T$ alone. For instance, for a given $T$ in the
diagonal basis of $S_{i}$, one may obtain $T_{d}$ from $\mathrm{Tr}\left[
T\right]  $ and $\det \left[  T\right]  $ and then solve Eq.~(\ref{wia}) for
mixing matrix elements.

In \cite{grimus2}, the two equations in (\ref{FG3}) are solved by employing a
theorem related to roots of unity \cite{CJ}. Here we note that substituting
Eq.~(\ref{lambdaij}) into (\ref{FG3}) leads to%
\begin{align}
\sum_{i=1}^{3}2\cos \left(  \frac{k^{(ij)}}{m^{(ij)}}2\pi \right)   &
=-2,\label{FG3T1}\\
\sum_{j=1}^{3}2\cos \left(  \frac{k^{(ij)}}{m^{(ij)}}2\pi \right)   &  =-2.
\label{FG3T2}%
\end{align}
In addition, in the minimal scenario the unitary condition in Eq.~(\ref{ucon0}%
) can also be written in a similar form, as shown in section~V. In the next
section, we introduce a method to handle trigonometric Diophantine equations,
such as the two equations above. Using this method, we derive the solutions to
Eqs.~(\ref{FG3T1}) and (\ref{FG3T2}) in Sec. IV.

\section{Trigonometric Diophantine equations}

In this section, we consider trigonometric Diophantine equations that can be
written as%
\begin{equation}
\sum \limits_{j=1}^{n}2\cos \frac{n_{j}}{p_{j}}2\pi=r \label{TDE0}%
\end{equation}
where $n_{j}$, $p_{j}$, and $r$ are rational integers. Without loss of
generality, we require that $n_{j}$ and $p_{j}$ are coprime or $\gcd
(n_{j},p_{j})=1$ where $\gcd$ stands for the greatest common divisor. To
simplify notations, below we denote rational angles such as $\left.  2\pi
n\right/  m$ by $\alpha_{nm}$ and $2\cos \alpha_{n_{j}p_{j}}$ by $\beta^{j}$.
Then Eq.~(\ref{TDE0}) can be written as%
\begin{equation}
\sum \limits_{j=1}^{n}\beta^{j}=\sum \limits_{j=1}^{n}2\cos \alpha_{n_{j}p_{j}%
}=r. \label{TDE1}%
\end{equation}
We will show that this equation can be transformed to a simple arithmetic
equation\footnote{Similar results may exist somewhere in literature, but we
are not aware of any of them.}, i.e.,%
\begin{equation}
\sum \limits_{j=1}^{n}\frac{s_{j}}{d_{j}}=r \label{TDE5}%
\end{equation}
where $d_{j}$ is the degree of the minimal polynomial (MP) of $\beta^{j}$,
i.e., the polynomial with integer coefficients satisfied by $\beta^{j}$ that
has the lowest degree\footnote{In other words, it is irreducible or cannot be
written as a product of two monic polynomials with integer coefficients. More
can be found in the appendix.}, and $s_{j}$ is the sum of all the roots of
that MP. Note that not only $d_{j}$, but also $s_{j}$, are rational integers,
as explained below. From solutions to Eq.~(\ref{TDE5}) one can find solutions
to Eq.~(\ref{TDE0}) because $p_{j}$ can be determined by $d_{j}$ and $s_{j}$.
To derive Eq.~(\ref{TDE5}), we will need some notions and results from
algebraic number theory, mostly for convenience. The detailed derivation is
presented below in a rather self-contained way, but, for conciseness, some
details are relegated to the appendix. Before proceed, we should mention that
in this paper we only use polynomials with integer coefficients, unless
otherwise stated. In addition, the set of all rational numbers and the set of
all rational integers are denoted by $\mathbb{Q}$ and $\mathbb{Z}$, respectively.

First, note that from
\[
\cos2\pi n_{j}=\cos \left(  p_{j}\cdot \alpha_{n_{j}p_{j}}\right)  =1
\]
and the expansion of $\cos \left(  p_{j}\alpha_{n_{j}p_{j}}\right)  $ in powers
of $\cos \left(  \alpha_{n_{j}p_{j}}\right)  $, as shown in the appendix, one
can see that $\beta^{j}=2\cos \alpha_{n_{j}p_{j}}$ are algebraic integers which
are defined to be solutions to monic polynomials with integer coefficients.
Moreover, all the $\beta^{j}=2\cos \left(  \left.  2\pi n_{j}\right/
p_{j}\right)  $ in Eq.~(\ref{TDE1}) can be expanded in powers of
\[
\gamma_{1}\equiv2\cos \frac{2\pi}{q}%
\]
where $q\equiv \mathrm{LCM}(p_{1},\ldots,p_{n})$ and $\mathrm{LCM}$ stands for
the least common multiple. Obviously, $\gamma_{1}$ is also an algebraic
integer. Let polynomial $g(x)$ of degree $d$ be the MP for $\gamma_{1}$. It
can be written as%
\[
g(x)=x^{d}+c_{d-1}x^{d-1}+\cdots+c_{0}=\prod_{i=1}^{d}(x-\gamma_{i})
\]
where $\gamma_{i}$ are the roots of $g(x)$. Since the degree of an algebraic
integer is defined to be the degree of its MP, we have $\deg(\gamma_{1})=d$
where $\deg(\gamma_{1})$ denotes the degree of $\gamma_{1}$.

To proceed, we notice that, using $g(\gamma_{1})=0$, one can eliminate any
term having a power of $\gamma_{1}$ higher than $d-1$ from a polynomial in
$\gamma_{1}$. Hence, if an algebraic integer $\theta$ can be written as a
polynomial in $\gamma_{1}$, then one can always write $\theta$ as
\begin{equation}
\theta=a_{0}+a_{1}\gamma_{1}+\cdots+a_{d-1}\gamma_{1}^{d-1}=\sum
\limits_{k=0}^{d-1}a_{k}\left(  \gamma_{1}\right)  ^{k}\equiv p(\gamma_{1})
\label{Qg1}%
\end{equation}
where $a_{i}\in \mathbb{Q}$. More importantly, the expression for $\theta$ in
the form of Eq.~(\ref{Qg1}) is unique. To show that, suppose $\theta$ can also
be written as%
\[
\theta=a_{0}^{\prime}+a_{1}^{\prime}\gamma_{1}+\cdots+a_{d-1}^{\prime}%
\gamma_{1}^{d-1}.
\]
Then one has%
\[
\left(  a_{d-1}^{\prime}-a_{d-1}\right)  \gamma_{1}^{d-1}+\cdots+\left(
a_{1}^{\prime}-a_{1}\right)  \gamma_{1}+\left(  a_{0}^{\prime}-a_{0}\right)
=0.
\]
Since the degree of $\gamma_{1}$ is $d$ and thus $\gamma_{1}$ cannot satisfy
any polynomial of degree less than $d$, it then follows that $a_{k}^{\prime
}=a_{k}$ for all $k$.

According to the discussion above, we can write $\beta^{j}$ uniquely as%
\begin{equation}
\beta^{j}=\sum \limits_{k=0}^{d-1}b_{k}^{j}\left(  \gamma_{1}\right)
^{k}\equiv \tilde{\beta}^{j}(\gamma_{1}) \label{betaj}%
\end{equation}
where $b_{k}\in \mathbb{Z}$. Then, from Eqs.~(\ref{TDE1}) and (\ref{betaj}) one
has%
\begin{equation}
\sum \limits_{j=1}^{n}\tilde{\beta}^{j}(\gamma_{1})=\sum \limits_{k=0}%
^{d-1}\left(  \sum \limits_{j=1}^{n}b_{k}^{j}\right)  \left(  \gamma
_{1}\right)  ^{k}=r. \label{TDE2}%
\end{equation}
Now comes a crucial step. Again, since $\gamma_{1}$ cannot satisfy any
polynomial of degree less than $d$, in Eq.~(\ref{TDE2}) one must have%
\[
\sum \limits_{j=1}^{n}b_{k}^{j}=0
\]
for $1\leq k\leq d-1$ because $r$ is a rational integer. Hence, in
Eq.~(\ref{TDE2}) one may replace $\gamma_{1}$ by any number. For our purpose,
we will substitute $\gamma_{1}$ by $\gamma_{2}$, $\ldots$, $\gamma_{d}$, i.e.,
the other roots of its MP. Then, besides Eq.~(\ref{TDE2}), we also have%
\[
\sum \limits_{j=1}^{n}\tilde{\beta}^{j}(\gamma_{i})=r
\]
for $2\leq i\leq d$. Summing over $i$ leads to
\begin{equation}
\sum \limits_{j=1}^{n}\left[  \sum \limits_{i=1}^{d}\tilde{\beta}^{j}(\gamma
_{i})\right]  =rd. \label{TDE3}%
\end{equation}
In addition, as discussed in the appendix, one can show that for any
$\gamma_{i}$ the corresponding $\tilde{\beta}^{j}(\gamma_{i})$ is a root of
the MP of $\tilde{\beta}^{j}(\gamma_{1})$ because $\gamma_{i}$ is a root of
the MP of $\gamma_{1}$. Now, let $d_{j}$ be the degree of $\beta^{j}%
=\tilde{\beta}^{j}(\gamma_{1})$. Because one can also show that when
$\gamma_{i}$ in $\tilde{\beta}^{j}(\gamma_{i})$ runs from $\gamma_{1}$ to
$\gamma_{d}$, each root of the MP of $\tilde{\beta}^{j}(\gamma_{1})$ repeats
$d/d_{j}$ times, we have%
\begin{equation}
\sum \limits_{i=1}^{d}\tilde{\beta}^{j}(\gamma_{i})=\frac{d}{d_{j}}s_{j}
\label{TDE4}%
\end{equation}
where $s_{j}$ is the sum of all the roots of the MP of $\beta^{j}$. From
Eqs.~(\ref{TDE3}) and (\ref{TDE4}) it follows that
\begin{equation}
\sum \limits_{j=1}^{n}\frac{s_{j}}{d_{j}}=\sum \limits_{j=1}^{n}\frac{1}%
{\deg \left(  \beta^{j}\right)  }\sum \nolimits_{c}\beta^{j}=r \label{TDE6}%
\end{equation}
where $\deg \left(  \beta^{j}\right)  =d_{j}$ and\quad$\sum \nolimits_{c}%
\beta^{j}=s_{j}$. We would like to emphasize that $\sum \nolimits_{c}\beta^{j}$
should not be confused with $\sum \nolimits_{j}\beta^{j}$. In addition, since
the MP of $\beta^{j}$ can be written as%
\[
x^{d_{j}}-s_{j}x^{d_{j}-1}+\cdots,
\]
then $s_{j}$ must be a rational integer.

In the appendix, the explicit expressions for $d_{m}=\deg \left(  2\cos
\alpha_{nm}\right)  $ and $s_{m}=\sum \nolimits_{c}2\cos \alpha_{nm}$ are given
in Eq.~(\ref{eulerp}) and Eq.~(\ref{csum2c}). As examples, $d_{m}$ and $s_{m}$
for $2\leq m\leq20$ are given in the table below, which will also be used in
the next two sections. Note that $d_{m}$ and $s_{m}$ together can determine
the value of $m$ in $\alpha_{nm}=2\pi n/m$ but not $n$ which can be any
integer that is relatively prime to $m$. For obvious reasons, we can require
that $0<n<m/2$. In addition, the value of $m$ may not be determined uniquely
by $d_{m}$ and $s_{m}$, as can be seen from table \ref{ephi}.

\begin{table}[h]
\centering
\begin{tabular}
[c]%
{|m{0.2in}|m{0.2in}|m{0.2in}|m{0.2in}|m{0.2in}|m{0.2in}|m{0.2in}|m{0.2in}|m{0.2in}|m{0.2in}|m{0.2in}|m{0.2in}|m{0.2in}|m{0.2in}|m{0.2in}|m{0.2in}|m{0.2in}|m{0.2in}|m{0.2in}|m{0.2in}|m{0.2in}|}%
\hline
$m$ & 1 & 2 & 3 & 4 & 5 & 6 & 7 & 8 & 9 & 10 & 11 & 12 & 13 & 14 & 15 & 16 &
17 & 18 & 19 & 20\\ \hline
$d_{m}$ & 1 & 1 & 1 & 1 & 2 & 1 & 3 & 2 & 3 & 2 & 5 & 2 & 6 & 3 & 4 & 4 & 8 &
3 & 9 & 4\\ \hline
$s_{m}$ & $2$ & $-2$ & $-1$ & $0$ & $-1$ & $1$ & $-1$ & $0$ & $0$ & $1$ & $-1$
& $0$ & $-1$ & $1$ & $1$ & $0$ & $-1$ & $0$ & $-1$ & $0$\\ \hline
\end{tabular}
\caption{The degree of the MP of $2 \cos \alpha_{nm}$ and the sum of its roots}%
\label{ephi}%
\end{table}

The issues mentioned above indicate that Eq.~(\ref{TDE6}) is only a necessary
condition for any set of $p_{j}$ or, more accurately, $\beta^{j}=2\cos \left(
\left.  2\pi n_{j}\right/  p_{j}\right)  $ to satisfy Eq.~(\ref{TDE0}).
Although it is not a sufficient condition and hence some solutions to
Eq.~(\ref{TDE6}) may not satisfy Eq.~(\ref{TDE0}), because it is not hard to
find its solutions in many cases, Eq.~(\ref{TDE6}) can be regarded as an
auxiliary condition which may greatly simplify the problem of finding
solutions to Eq.~(\ref{TDE0}). As to $n_{j}$ in Eq.~(\ref{TDE0}), one can find
them by trial and error, especially when $d_{j}$ is small because, as
explained in the appendix, $d_{j}$ is the number of $n_{j}$ satisfying
$\gcd(n_{j},p_{j})=1$ and $0<n_{j}<p_{j}/2$.

Eq.~(\ref{TDE6}) is most helpful in the cases where it can provide enough
information for us to solve Eq.~(\ref{TDE0}) completely. Such situation may
occur if the number of terms on the left-hand side or $n$ is not large and
$|r|$ is comparable to $n$. The reason is that $s_{j}=\sum \nolimits_{c}%
2\cos \left(  \left.  2\pi n_{j}\right/  p_{j}\right)  $ can only be $\pm1$ or
$0$ for all $p_{j}$ (except for $p_{j}=1$ or $2$) and the lower bound on
$d_{j}$ increases with $p_{j}$ (see the discussion below Eq.~[\ref{eulerp}] in
the appendix). Hence, for a small $n$ and a $|r|\sim n$, to satisfy
Eq.~(\ref{TDE6}), $p_{j}$ cannot all be very large in most cases and thus it
might be possible to solve it completely. This is exactly the situation for
the two cases discussed in the following two sections. This method may also be
useful in some other cases, for example, when one only wants to find certain
specific solutions or for some reason $p_{j}$ can only be chosen from a given
set of numbers.

\section{Residual symmetry in the charged lepton sector}

In this section, we will show that the complete set of solutions to
Eqs.~(\ref{FG3T1}) and (\ref{FG3T2}) can be easily obtained by the method
introduced above. Since these two equations are similar, only Eq.~(\ref{FG3T1}%
) will be discussed in detail. As above, to simplify notations, we write
Eq.~(\ref{FG3T1}) as%
\begin{equation}
\sum_{i=1}^{3}\beta^{(ij)}=\sum_{i=1}^{3}2\cos \alpha^{(ij)}=-2. \label{FG3Ta}%
\end{equation}
where $\alpha^{(ij)}=\left.  2\pi k^{(ij)}\right/  m^{(ij)}$. Then, as shown
in the previous section, this equation can be transformed to%
\begin{equation}
\sum \limits_{i=1}^{3}y_{i}=\sum \limits_{i=1}^{3}\frac{s_{i}}{d_{i}}%
=\sum \limits_{i=1}^{3}\frac{1}{\deg \left(  \beta^{(ij)}\right)  }%
\sum \nolimits_{c}\beta^{(ij)}=-2 \label{FG3Tb}%
\end{equation}
where $y_{i}=s_{i}/d_{i}$, $s_{i}=\sum \nolimits_{c}\beta^{(ij)}$, and
$d_{i}=\deg \left(  \beta^{(ij)}\right)  $.

Since, as discussed above, $\left \vert s_{i}\right \vert \leq1$ for
$m^{(ij)}>2$ and $d_{i}$ tend to grow with $m^{(ij)}$, one can show that%
\begin{equation}
\left \vert \frac{s_{j}}{d_{j}}\right \vert \leq2 \label{sdineq}%
\end{equation}
for any $j$. Hence, without loss of generality, we can require that
\begin{equation}
2\geq|y_{1}|\geq|y_{2}|\geq|y_{3}|. \label{y123ineq}%
\end{equation}
Then from Eq.~(\ref{FG3Tb}) it follows that%
\[
|y_{1}|=\left \vert \frac{s_{1}}{d_{1}}\right \vert \geq \frac{2}{3}%
\]
and hence in table \ref{ephi} one finds that
\[
m^{(1j)}=1,\ 2,\ 3,\  \text{or\ }\;6.
\]

If $m^{(1j)}=1$, then
\begin{equation}
\alpha^{(1j)}=2\pi,\quad \beta^{(1j)}=2\cos \alpha^{(1j)}=2. \label{FG3Taj1a}%
\end{equation}
Substituting $\beta^{(1j)}=2$ into Eq.~(\ref{FG3Ta}) leads to%
\[
2\cos \alpha^{(2j)}+2\cos \alpha^{(3j)}=-4.
\]
Therefore, $m^{(2j)}=m^{(3j)}=2$ and hence%
\begin{equation}
\beta^{(2j)}=\beta^{(3j)}=2\cos \left(  \frac{1}{2}\times2\pi \right)  =-2.
\label{FG3Taj1b}%
\end{equation}

In the case where $m^{(1j)}=2$, one has $\alpha^{(1j)}=\pi$ and%
\begin{equation}
\beta^{(1j)}=2\cos \alpha^{(1j)}=-2. \label{FG3Taj2a}%
\end{equation}
Then from Eq.~(\ref{FG3Ta}) it follows that%
\begin{equation}
\beta^{(2j)}=-\beta^{(3j)}=2\cos \varphi \label{FG3Taj2b}%
\end{equation}
where $\varphi=\alpha^{(2j)}$ is an arbitrary rational angle. Note that, no
further constraint can be imposed on $\varphi$ in this case because
Eq.~(\ref{FG3Taj2b}) can be satisfied if
\[
\frac{k^{(2j)}}{m^{(2j)}}+\frac{k^{(3j)}}{m^{(3j)}}=\frac{1}{2}%
\]
from which one can find a $m^{(3j)}$ for any $m^{(2j)}$.

The next case is the one where $m^{(1j)}=3$, which is a bit more complicated.
From table \ref{ephi}, one finds that $d_{1}=1$ and $s_{1}=-1$ and hence
$y_{1}=-1$. Then from Eq.~(\ref{FG3Tb}), one has $y_{2}+y_{3}=-1$. Together
with Eq.~(\ref{y123ineq}) and $|y_{1}|=1$, this leads to $1/2\leq|y_{2}|\leq1$
and $|y_{3}|\leq1$. From these results one finds that $-1\leq y_{2}\leq-1/2$.
Therefore, $y_{2}=-1$ or $-1/2$, as can be seen from table \ref{ephi}. If
$y_{2}=-1$, one has $m^{(2j)}=3$. Together with $m^{(1j)}=3$ and
Eq.~(\ref{FG3Ta}), it leads to $\beta^{(3j)}=0$ and hence $m^{(3j)}=4$. In
short, if $y_{2}=-1$, one has%
\begin{equation}%
\begin{array}
[c]{l}%
\displaystyle \beta^{(1j)}=\beta^{(2j)}=2\cos \frac{2\pi}{3}=-1,\medskip \\
\displaystyle \beta^{(3j)}=2\cos \frac{2\pi}{4}=0.
\end{array}
\label{FG3Taj3}%
\end{equation}
If $y_{2}=-1/2$, then $y_{3}=-1/2$. From table \ref{ephi}, one has
$m^{(2j)}=m^{(3j)}=5$. Because $d_{2}=d_{3}=2$, in Eq.~(\ref{FG3Ta}) both
$k^{(2j)}$ and $k^{(3j)}$ have two choices. The solution to Eq.~(\ref{FG3Ta})
is found to be $k^{(2j)}=1$ and $k^{(3j)}=2$. Therefore, one has%
\begin{equation}%
\begin{array}
[c]{l}%
\displaystyle \beta^{(1j)}=2\cos \frac{2\pi}{3}=-1,\medskip \\
\displaystyle \beta^{(2j)}=2\cos \frac{2\pi}{5},\quad \beta^{(3j)}=2\cos
\frac{4\pi}{5}.
\end{array}
\label{FG3Taj4}%
\end{equation}

After that, we are left with the case where $m^{(1j)}=6$. In table \ref{ephi}
one finds that $d_{1}=1$ and $s_{1}=1$. However, from Eq.~(\ref{FG3Tb}) it
follows that $y_{2}+y_{3}=-3$ which is not consistent with Eq.~(\ref{y123ineq}%
), since the latter would lead to $|y_{2}+y_{3}|\leq2|y_{1}|=2$. Therefore, no
solution satisfying Eq.~(\ref{y123ineq}) exists for $m^{(1j)}=6$.

Now, we have found all the solutions to Eq.~(\ref{FG3Ta}) which are given by
Eqs.~(\ref{FG3Taj1a}-\ref{FG3Taj4}). Since, from Eqs.~(\ref{lambdaij}) and
(\ref{TijAbs2}) one has%
\begin{equation}
4\left \vert T_{ij}\right \vert ^{2}=\beta^{(ij)}+2=2\cos \left(  \frac{k^{(ij)}%
}{m^{(ij)}}2\pi \right)  +2, \label{TijAbs2a}%
\end{equation}
then, for each solution, the corresponding $\left \vert T_{ij}\right \vert ^{2}$
can be obtained from the relation above and written collectively as $\left(
\left \vert T_{1j}\right \vert ^{2},\left \vert T_{2j}\right \vert ^{2},\left \vert
T_{3j}\right \vert ^{2}\right)  $. For the four solutions found above, one has
the following four possibilities:%
\begin{align*}
&  \left(  1,0,0\right)  ,\quad \left(  \frac{1}{4},\frac{3+\sqrt{5}}{8}%
,\frac{3-\sqrt{5}}{8}\right)  ,\\
&  \left(  \frac{1}{2},\frac{1}{4},\frac{1}{4}\right)  ,\quad \left(
0,\frac{1+\cos \varphi}{2},\frac{1-\cos \varphi}{2}\right)  .
\end{align*}
Note that the order of $\left \vert T_{ij}\right \vert ^{2}$ can be changed
because in Eq.~(\ref{y123ineq}) the order of $y_{1}$, $y_{2}$, and $y_{3}$ can
also be changed, which may lead to solutions with different orderings.

It is easy to see that the solutions to the second equation in (\ref{FG3}) are
the same as the first one. Hence, in matrix $|T|^{2}$ where $\left(
|T|^{2}\right)  _{ij}=\left \vert T_{ij}\right \vert ^{2}$ every row and column
must be one of the four possibilities given above. It is then not hard to find
that there exist only five basic forms that $|T|$ can assume, as shown in
\cite{grimus2}. After that, by a thorough and careful analysis, which can also
be found in \cite{grimus2}, one can find all the possible full-mixing patterns.

Before proceeding to the next section, we would like to mention that some
results used in our discussions might also be useful on other occasions. For
example, because a non-integer rational number is not an algebraic integer, as
shown in the appendix, sometimes this can provide a quick way to show that for
a given complex number $\eta$ there does not exist a rational integer $m$
satisfying $\eta^{m}=1$. If such a $m$ exists, then $\eta$ is a root of unity
and can be written as $\eta=e^{i2\pi \left.  k\right/  m}$. Hence $\eta
+\eta^{\ast}=2\cos \left(  2\pi \left.  k\right/  m\right)  $ must be an
algebraic integer. But, if $\eta+\eta^{\ast}$ is a non-integer rational
number, then it cannot be an algebraic integer and thus $\eta$ cannot be a
root of unity. For instance, $\eta=\left.  \left(  1+i3\sqrt{7}\right)
\right/  8$ or $\eta=\left.  \left(  -1+i\sqrt{15}\right)  \right/  4$ cannot
be a root of unity because $\eta+\eta^{\ast}$ is a non-integer rational
number\footnote{These two examples are taken from \cite{grimus2} where a
somewhat different argument is used.}.

\section{Mixing patterns in the minimal scenario}

The derivation of lepton mixing patterns from $|T|$ is rather complicated. In
this section, we will show that, in the minimal scenario where both
$\mathrm{Tr}[T]$ and $\mathrm{Tr}[W_{i}]$ are real, the unitarity condition
(\ref{ucon0}) can also be solved completely and hence full-mixing matrices can
be obtained in a rather straightforward way.

When $\mathrm{Tr}[T]$ is real and $T$ belongs to $SU(3)$, from Eq.~(\ref{ba})
it follows that $T_{d}$ can be written as one of the following three matrices:%
\begin{align}
T_{e}  &  \equiv T_{1}=\mathrm{diag}\{1,e^{2\pi in_{4}/p_{4}},e^{-2\pi
in_{4}/p_{4}}\},\nonumber \\
T_{\mu}  &  \equiv T_{2}=\mathrm{diag}\{e^{2\pi in_{4}/p_{4}},1,e^{-2\pi
in_{4}/p_{4}}\},\label{ti}\\
T_{\tau}  &  \equiv T_{3}=\mathrm{diag}\{e^{2\pi in_{4}/p_{4}},e^{-2\pi
in_{4}/p_{4}},1\}.\nonumber
\end{align}
where $n_{4}$ and $p_{4}$ are coprime. It then follows that%
\begin{equation}
\mathrm{Tr}[T]=1+2\cos2\pi n_{4}/p_{4}. \label{trt}%
\end{equation}
In this section, we will denote $T_{d}$ by $T_{\alpha}$ where $\alpha=1$, $2$,
or $3$. In addition, we require that $m>2$, because otherwise the lepton
mixing matrix cannot be fully determined. Similarly, if $\mathrm{Tr}[W_{i}]$
is real, because $\left(  W_{i}\right)  ^{p_{i}}=\mathbb{I}$, the eigenvalues
of $W_{i}$ can always be written as $1$, $e^{2\pi in_{i}/p_{i}}$, or $e^{-2\pi
in_{i}/p_{i}}$ with $\gcd(n_{i},p_{i})=1$. Hence,%
\begin{equation}
\mathrm{Tr}[W_{i}]=1+2\cos2\pi n_{i}/p_{i}. \label{twia}%
\end{equation}
Then, from Eq.~(\ref{wia}), Eq.~(\ref{trt}), Eq.~(\ref{twia}), and $\sum
_{\rho=1}^{3}|U_{\rho i}|^{2}=1$, one finds that%
\begin{align}
\left \vert U_{\alpha i}\right \vert ^{2}  &  =\frac{1+\mathrm{Tr}[W_{i}%
]}{3-\mathrm{Tr}[T_{\alpha}]}=\frac{1+\cos \left(  2\pi n_{i}/p_{i}\right)
}{2\sin^{2}\left(  \pi n_{4}/p_{4}\right)  },\label{meq3}\\
\left \vert U_{\beta i}\right \vert ^{2}  &  =\left \vert U_{\gamma i}\right \vert
^{2}=\frac{1}{2}\left(  1-|U_{\alpha i}|^{2}\right)  , \label{meq4}%
\end{align}
where $\beta,\gamma \neq \alpha$ and $\beta<\gamma$. For detailed derivation,
see \cite{Sim1}.

As discussed in Sec.~II, to generate a full-mixing matrix, $W_{i}$ and
$T_{\alpha}$ must satisfy the unitary condition in Eq.~(\ref{ucon0}). In the
minimal scenario, substituting Eqs.~(\ref{trt}) and (\ref{twia}) into
Eq.~(\ref{ucon0}) leads to%
\begin{equation}
\sum_{j=1}^{4}\beta^{j}=\sum_{j=1}^{4}2\cos \alpha_{n_{j}p_{j}}=-4 \label{ucon}%
\end{equation}
where $\beta^{j}=2\cos \alpha_{n_{j}p_{j}}$ and $\alpha_{n_{j}p_{j}}=\left.
2\pi n_{j}\right/  p_{j}$. As in the previous section, this equation can be
transformed to%
\begin{equation}
\sum_{j=1}^{4}y_{j}=\sum_{j=1}^{4}\frac{s_{j}}{d_{j}}=\sum_{j=1}^{4}\frac
{1}{\deg \left(  \beta^{j}\right)  }\sum \nolimits_{c}\beta^{j}=-4 \label{deq1}%
\end{equation}
which can also be easily solved in a way similar to the previous section.

First, note that Eq.~(\ref{deq1}) cannot be satisfied if all $y_{j}>-1$, then
we can require that $y_{1}\leq-1$ and $y_{1}\leq y_{j}$ for $j\geq2$. From
table \ref{ephi} one finds that $y_{1}=-1$ or $-2$ corresponding to $d_{1}=1$
and $s_{1}=-1$ or $-2$.

If $d_{1}=1$ and $s_{1}=-1$, from Eq.~(\ref{deq1}) it follows that $\sum
_{j=2}^{4}y_{j}=-3$. Since $y_{j}\geq y_{1}=-1$ for $j\geq2$, it cannot be
satisfied unless all $y_{j}=-1$. Therefore, from table \ref{ephi} one finds
that $s_{j}=-1$ and $d_{j}=1$ and hence $p_{j}=3$ for all $j$.

If $d_{1}=1$ and $s_{1}=-2$, from table \ref{ephi} one finds that $p_{1}=2$.
Since one can only let $n_{1}=1$, from Eq.~(\ref{ucon}) it follows that
$\sum_{j=2}^{4}\beta^{j}=-2$. Besides that, from Eq.~(\ref{deq1}) and
$y_{1}=-2$ one has $\sum_{j=2}^{4}y_{j}=-2$. These equations are identical to
Eqs.~(\ref{FG3Ta}) and (\ref{FG3Tb}) and hence their solutions can be obtained
from those found in Sec.~IV.

Here we denote the solutions to Eq.~(\ref{ucon}) by $\{p_{1},p_{2},p_{3}%
,p_{4}\}$ from which the corresponding mixing matrix can be obtained from
Eqs.~(\ref{meq3}) and (\ref{meq4}). In this notation the solutions we found
can be written as $\{3,3,3,3\}$, $\{2,1,2,2\}$, $\{2,2,p_{3},p_{4}\}$,
$\{2,3,3,4\}$ and $\{2,3,5,5\}$. As explained in the previous section, in the
third solution, $p_{3}$ and $p_{4}$ can be any integers that satisfy the
relation $n_{3}/p_{3}+n_{4}/p_{4}=1/2$ where $n_{3}$ and $n_{4}$ are arbitrary
integers satisfying $0<n_{j}/p_{j}\leq1/2$ and $\gcd(n_{j},p_{j})=1$. In
Eq.~(\ref{deq1}) one may change the order of $y_{j}$ and hence the order of
$p_{j}$ can also be changed in these solutions.\ Nevertheless, as discussed in
\cite{BH}, only the last two solutions can lead to phenomenologically
interesting mixing patterns including TBM, BM, or the golden ratio mixings.
For more discussion, see \cite{BH}.

\section{Discussions}

In the previous two sections, under the assumption that residual symmetries
are originated from a finite group, we discussed the condition constraining
the symmetry of the charged lepton mass matrix in the basis where the neutrino
mass matrix is diagonal and the unitary condition in the minimal scenario.
Both of them can be completely solved by the algebraic method presented in
section III and their solutions can be used to construct full-mixing matrices.
This method can also be used to solve trigonometric Diophantine equations
similar to those discussed in this paper.

It is interesting to see that, besides group theoretical techniques, other
algebraic techniques can also be useful and even the key to some important
results including the basic forms of $|T|$ discussed in section IV. As another
example, in \cite{grimus2} it was found that, under the same assumptions
adopted in this paper, only particular trimaximal mixing patterns can survive
the current neutrino oscillation data. The corresponding symmetry groups
include, e.g., $\mathbf{\Delta}(600)$, $\mathbf{\Delta}(1536)$, and
$(\mathbf{Z}_{18}\times \mathbf{Z}_{6})\rtimes \mathbf{S}_{3}$, which can also
be obtained from solutions to Eq.~(\ref{ucon0}) in non-minimal scenarios. To
find those solutions, notions and tools from algebraic number theory can also
be helpful, as discussed in \cite{BH}. In addition, from Eq.~(\ref{wisum}) one
finds that the unitary condition in Eq.~(\ref{ucon0}) can be written as a
vanishing sum of $12$ roots of unity. The classification of its solutions can
be found in \cite{Poonen} from which one may find another approach to the
classification of full-mixing patterns. Hopefully, the works presented in some
papers including \cite{grimus1}, \cite{grimus2}, \cite{BH}, and this one may
draw attention to those less well-known techniques.

\textit{Acknowledgements} -- This work was supported in part by the National
Science Foundation of China (NSFC) under the grant 10965003. It was also
partly supported by 555 talent project of Jiangxi Province.

\appendix

\section{Trigonometric algebraic numbers}

In this appendix, some mathematical details concerning trigonometric algebraic
numbers will be provided. Below we first quickly recall some basic notions
from algebraic number theory \cite{poll}:

\begin{enumerate}
\item An algebraic number $\theta$ over $\mathbb{Q}$, the set (or field) of
all rational numbers, is a root of a monic polynomial over $\mathbb{Q}$ which
can be written as
\begin{equation}
f(x)=x^{n}+a_{n-1}x^{n-1}+\ldots+a_{0} \label{ai1}%
\end{equation}
where all $a_{i}\in \mathbb{Q}$ and $n$ is called the degree of the polynomial.
Moreover, if all $a_{i}$ are rational integers, $\theta$ can also be called an
algebraic integer. If $\alpha$ and $\beta$ are algebraic numbers (integers),
then $\alpha+\beta$ and $\alpha \beta$ are also algebraic numbers (integers).
\end{enumerate}

Note that, unless otherwise mentioned, here we consider only algebraic numbers
(integers) and polynomials over $\mathbb{Q}$, as defined above. Hence, from
now on we will omit the phrase ``over $\mathbb{Q}$'' for the sake of conciseness.

\begin{enumerate}
\item[2.] The minimal polynomial (MP) of $\theta$ is the polynomial of lowest
degree that $\theta$ satisfies. One can show that the MP of $\theta$ is
unique, otherwise one can construct a polynomial satisfied by $\theta$ of
lower degree than the MP. For the same reason, the MP is clearly irreducible
or cannot be written as a product of two polynomials.

\item[3.] The degree of $\theta$, denoted by $\deg(\theta)$, is defined to be
the degree of its MP. All the roots of its MP are also called the conjugates
of $\theta$, which are denoted by $\theta_{k}$ where $k=1,2,\ldots,\deg
(\theta)$ and $\theta_{1}=\theta$. An important and frequently used result is
that any polynomial satisfied by $\theta$ must be divisible by its MP and
hence satisfied by all $\theta_{k}$ because the remainder of the division, if
not vanishing, is also satisfied by $\theta$, but of lower degree than its MP.
As a consequence, the MP of $\theta$ is also the MP of $\theta_{k}$ since it
is not reducible.

\item[4.] Since the MP of an algebraic number (integer) $\theta$ has rational
(integer) coefficients and can always be written as $\prod \nolimits_{k}%
(x-\theta_{k})$, it is obvious that $\sum \nolimits_{k}\theta_{k}$ is a
rational number (integer). One can also show that $\theta_{k}$ are distinct
because, if $\theta_{i}=\theta_{j}$, then from the MP one can construct a
polynomial satisfied by $\theta_{i}$ or $\theta_{j}$ of lower degree than the
MP by taking derivative with respect to $x$.
\end{enumerate}

Mostly for convenience, simple algebraic extension of $\mathbb{Q}$ is
introduced below. No deep result concerning field extensions is needed for our discussions.

\begin{enumerate}
\item[5.] By adjoining to $\mathbb{Q}$ an algebraic number $\theta$, the field
of rational numbers can be extended to another field denoted by $\mathbb{Q}%
(\theta)$. As shown in Sec.~III, every element $\lambda$ of $\mathbb{Q}%
(\theta)$ can be written uniquely in the form%
\[
\lambda=a_{0}+a_{1}\theta+\cdots+a_{v-1}\theta^{d-1}\equiv p(\theta)
\]
where $a_{i}\in \mathbb{Q}$ and $d=\deg(\theta)$. The conjugates of $\lambda$
\emph{for} $\mathbb{Q}(\theta)$ are defined to be $p(\theta_{k})\equiv
\bar{\lambda}_{k}$ where $\theta_{k}$ are the conjugates of $\theta$ and
$\bar{\lambda}_{1}=\lambda$.
\end{enumerate}

We should emphasize that $\bar{\lambda}_{i}$ defined above are the conjugates
of $\lambda$ \emph{for} $\mathbb{Q}(\theta)$, but not the conjugates of
$\lambda$ \textit{over} $\mathbb{Q}$ (defined in item 3), which are denoted by
$\lambda_{k}$. Nevertheless, they are closely related:

\begin{enumerate}
\item[6.] The set $\left \{  \bar{\lambda}_{1},\bar{\lambda}_{2},\ldots
,\bar{\lambda}_{d}\right \}  $ is not necessarily identical to the set of the
conjugates of $\lambda$ \emph{over} $\mathbb{Q}$, i.e., $\left \{  \lambda
_{1},\lambda_{2},\ldots,\lambda_{\deg(\lambda)}\right \}  $ where $\lambda
_{1}=\lambda=\bar{\lambda}_{1}$. However, one can show that in $\left \{
\bar{\lambda}_{1},\bar{\lambda}_{2},\ldots,\bar{\lambda}_{d}\right \}  $ each
$\lambda_{k}$ repeats $d\left/  \deg(\lambda)\right.  $ times.
\end{enumerate}

It might be instructive to go over the proof of the result given in item 6,
which is very important in the derivation of Eq.~(\ref{TDE5}). Below we follow
\cite{poll}. To begin with, one consider the polynomial\footnote{One can show
that the coefficients of $f(x)$ are rational numbers or integers because they
can be written as polynomials of the elementary symmetric functions in
$\theta_{1},\ldots \theta_{k}$ such as $\sum \nolimits_{k}\theta_{k}$,
$\sum \nolimits_{k,j}\theta_{k}\theta_{j}$, etc., which can be shown to be
rational numbers or integers since they are the coefficients of the MP of
$\theta_{k}$, i.e., $\prod \nolimits_{k}(x-\theta_{k})$. A complete proof of
this point would take us too far afield. One may find the details in
\cite{poll}.}%
\[
f(x)=\prod_{i=1}^{d}\left(  x-\bar{\lambda}_{i}\right)  =\prod_{i=1}%
^{d}\left(  x-p(\theta_{i})\right)  .
\]
Let the MP of $\lambda$ be $g(x)$ of degree $\deg(\lambda)$. Since
$\lambda=\bar{\lambda}_{1}$ is a root of $f(x)$, then, for the reason
explained in item 3, $f(x)$ must be divisible by $g(x)$. Therefore, one can
write%
\[
f(x)=\left[  g(x)\right]  ^{n}h(x)
\]
where $g(x)\nmid h(x)$. Then one can show that $h(x)$ must be a constant.
Otherwise, for some $\bar{\lambda}_{m}$, one must have $h\left(  \bar{\lambda
}_{m}\right)  =h\left(  p(\theta_{m})\right)  =0$. Since $\theta_{m}$
satisfies $h\left(  p(x)\right)  $, then $h\left(  p(x)\right)  $ must also be
satisfied by $\theta_{1}$ (see item 3). Therefore, one has $h\left(
p(\theta_{1})\right)  =h\left(  \bar{\lambda}_{1}\right)  =h\left(
\lambda \right)  =0$ from which it follows that $g(x)|h(x)$ which contradicts
the requirement $g(x)\nmid h(x)$. After that, it is not hard to see that
$h(x)=1$ and hence $f(x)=\left[  g(x)\right]  ^{n}$ and $n=d\left/
\deg(\lambda)\right.  $.

Below we will concentrate on trigonometric algebraic numbers, especially the
cosines of rational angles, i.e., $2\cos \alpha_{nm}$ where $\alpha_{nm}%
\equiv2\pi n/m$ and $\gcd(n,m)=1$. As in the main text, below we denote
$2\cos \alpha_{nm}$ by $\beta_{nm}$.

First of all, recall that \cite{Gradshteyn}
\[
2\cos m\alpha=\left(  2\cos \alpha \right)  ^{m}-m\left(  2\cos \alpha \right)
^{m-2}+\frac{m(m-3)}{2}\left(  2\cos \alpha \right)  ^{m-4}+\cdots.
\]
Replacing $2\cos \alpha$ by $x$ in the expression on the right-hand side, we
can define polynomials $c_{m}(x)$ as\footnote{If more details are needed, note
that $c_{m}(x)$ can also be written as $2T_{n}(x/2)-2$ where $T_{n}(x)$ are
Chebyshev polynomials of the first kind \cite{Gradshteyn}.}
\[
c_{m}(x)\equiv \left[  x^{m}-mx^{m-2}+\frac{m(m-3)}{2}x^{m-4}+\cdots \right]
-2.
\]
From $\cos m\alpha_{nm}=1$, one has%
\[
c_{m}(\beta_{nm})=2\cos m\alpha_{nm}-2=0
\]
from which it follows that $\beta_{nm}\equiv2\cos \alpha_{nm}$ is an algebraic
integer. An immediate consequence is that, if $\beta_{nm}$ is a rational
number, it must be a rational integer. Note that its MP is not $c_{m}(x)$, but
a factor of $c_{m}(x)$, as discussed above in item 3. Hence, any conjugate of
$\beta_{nm}$ must also be a root of $c_{m}(x)$.

Next, except for $m=1$ or $2$, the algebraic degree of $\beta_{nm}$ is given
by $\varphi(m)/2$ where $\varphi(m)$ is Euler's $\varphi$-function
\cite{hardy}, i.e.,
\begin{equation}
\deg(2\cos \alpha_{nm})=\frac{\varphi(m)}{2}=\frac{m}{2}\prod \limits_{P|m}%
(1-P^{-1}) \label{eulerp}%
\end{equation}
where $P$ are prime numbers and the product involves all the \emph{distinct}
prime factors of $m$. Recall that, by definition, $\varphi(m)$ is equal to the
the number of positive integers that are less than $m$ and are relatively
prime to $m$. A useful property of $\varphi(m)$ is that $\varphi(m)$ tends to
increase with $m$. In fact, various lower bounds on $\varphi(m)$ can be
established \cite{hont}. For instance, $\varphi(m)\geq \sqrt{m}$ for $m\neq2$
and $m\neq6$ and $\varphi(m)>m^{2/3}$ for $m>30$.

A rigorous proof of Eq.~(\ref{eulerp}) can be found in \cite{niven}. A short
but less elementary proof can be found in \cite{wat}. Some examples are given
in table \ref{ephi} of section~III. Below we give a simple argument showing
that the conjugates of $\beta_{nm}$, i.e., the roots of its MP, can include
only those $2\cos \alpha_{qm}$ with $\gcd(q,m)=1$, the number of which is
exactly $\varphi(m)/2$ after the equality between $2\cos \alpha_{qm}$ and
$2\cos \alpha_{\left(  m-q\right)  m}$ is taken into account. To show that, let
us consider $\beta_{kp}=2\cos \alpha_{kp}$ and $\beta_{nm}=2\cos \alpha_{nm}$
with $\gcd(k,p)=\gcd(n,m)=1$ and $p<m$. From $c_{p}(\beta_{kp})=0$ and%
\[
c_{p}(\beta_{nm})=2\cos p\alpha_{nm}-2\neq0
\]
one finds that $\beta_{nm}$ cannot be a root of $c_{p}(x)$ which contains the
MP of $\beta_{kp}$. Therefore, $\beta_{nm}$ is not a root of the MP of any
$2\cos \alpha_{qm}$ if $q$ is \textit{not} relatively prime to $m$. In other
words, the former is not a conjugate of the latter. Since all the conjugates
of $\beta_{nm}$ are among the roots of $c_{m}(x)$ which can be written as
$2\cos \alpha_{qm}$ with $1\leq q\leq m$, we are left with $2\cos \alpha_{qm}$
with $\gcd(q,m)=1$, as promised.

Now we turn to the sum of the conjugates of $\beta_{nm}$, including all the
distinct $2\cos \alpha_{qm}$ with $\gcd(q,m)=1$. The results are given by%
\begin{equation}
\sum \nolimits_{c}2\cos \alpha_{nm}=\left \{
\begin{array}
[c]{ll}%
2,\quad & m=1\\
-2, & m=2\\
\mu(m),\quad & m\geq3
\end{array}
\right.  \label{csum2c}%
\end{equation}
where $\sum \nolimits_{c}\theta$ denotes the sum of the conjugates of an
algebraic number $\theta$ and $\mu(m)$ is the M\"{o}bius function \cite{hardy}
defined as%
\[
\mu(m)=\left \{
\begin{array}
[c]{ll}%
1, & m=1\\
0, & \mathrm{if\ }m\mathrm{\ has\ a\ square\ factor}\\
(-1)^{k},\quad & \mathrm{if\ }m=p_{1}p_{2}\ldots p_{k}\mathrm{\ with\ }%
p_{i}\mathrm{\ being\ different\ prime\ numbers}%
\end{array}
\right.
\]
For some examples, see table \ref{ephi} in section III. Below we give an
example to demonstrate how Eq.~(\ref{csum2c}) is derived. First, one has%
\begin{equation}
\sum_{k=1}^{m-1}\cos \frac{k}{m}2\pi=\operatorname{Re}\left[  \sum_{k=1}%
^{m-1}e^{i\frac{k}{m}2\pi}\right]  =-1 \label{cossum}%
\end{equation}
for any $m\geq2$. Then one can divide the terms on the left-hand side into
groups of conjugates. For example, for $m=3\times7$,%
\begin{equation}
\sum_{k=1}^{20}\cos \frac{k}{21}2\pi=\sum_{i=1}^{2}\cos \frac{i}{3}2\pi
+\sum_{i=1}^{6}\cos \frac{i}{7}2\pi+2\sum_{\substack{j\leq10,\\(j,21)=1}%
}\cos \frac{j}{21}2\pi \label{cossum21}%
\end{equation}
where $(j,21)$ is short for $\gcd(j,21)$. As explained above, the last term
can be written as $\sum \nolimits_{c}2\cos \left(  2\pi n/21\right)  $ where $n$
can be any integer satisfying $\gcd(n,21)=1$. From Eq.~(\ref{cossum}) and
(\ref{cossum21}) one finds that%
\[
\sum \nolimits_{c}2\cos \frac{n}{21}2\pi=1.
\]
By induction the result given by Eq.~(\ref{csum2c}) can also be proved for
arbitrary $m$ in a similar way. The detailed derivation is somewhat lengthy
and hence will not be presented here. In addition, for a result like
Eq.~(\ref{csum2c}), it seems to us that numerical verification might be a
simple way to convince oneself. We have done that for $m\leq5\times10^{4}$,
which should be sufficient for us because in our discussions $m$ represents
the order of a group element.

\end{document}